\documentclass{moriond}

\def\be{\begin{equation}}
\def\ee{\end{equation}}
\def\bea{\begin{eqnarray}}
\def\eea{\end{eqnarray}}

\usepackage{comment}
\usepackage{hyperref}

\newcommand{\Photo}{\includegraphics[height=35mm]{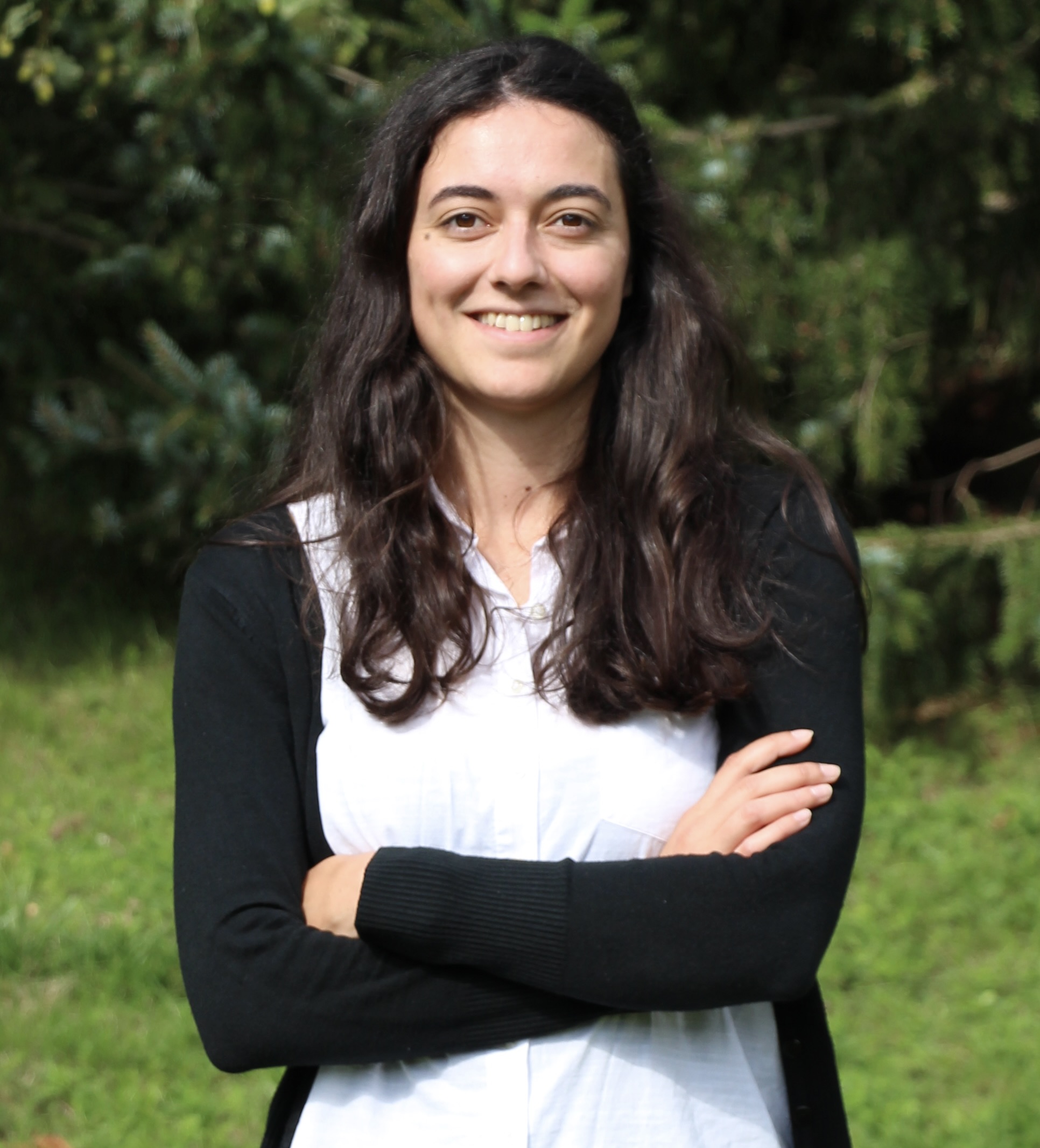}}

\begin{document}
\vspace*{4cm}
\title{Flavour changing charged current decays at LHCb}

\author{Biljana Mitreska on behalf of the LHCb collaboration}

\address{Department of Physics and Astronomy, University of Manchester,\\ Oxford Road M13 9PL, Manchester United Kingdom}

\maketitle\abstracts{Semileptonic $b$-hadron decays proceed via charged-current interactions and provide powerful probes for testing the Standard Model and searching for New Physics effects. The advantages of studying such decays include the large branching fractions and reliable calculations of the hadronic matrix elements. Several features can be studied, such as the ratios of branching fractions, CKM parameters, properties of $b$-hadron production, form factor parameters and New Physics Wilson coefficients. In this contribution, LHCb measurements of branching fraction in $\Lambda \to p \mu^{-} \bar{\nu}_{\mu}$ and form factor parameters with $B^0 \to D^{*+} \mu^{-} \nu_{\mu}$ decays are presented.}

\section{Introduction}
\begin{figure}[h]
    \centering
    \includegraphics[width=0.4\textwidth]{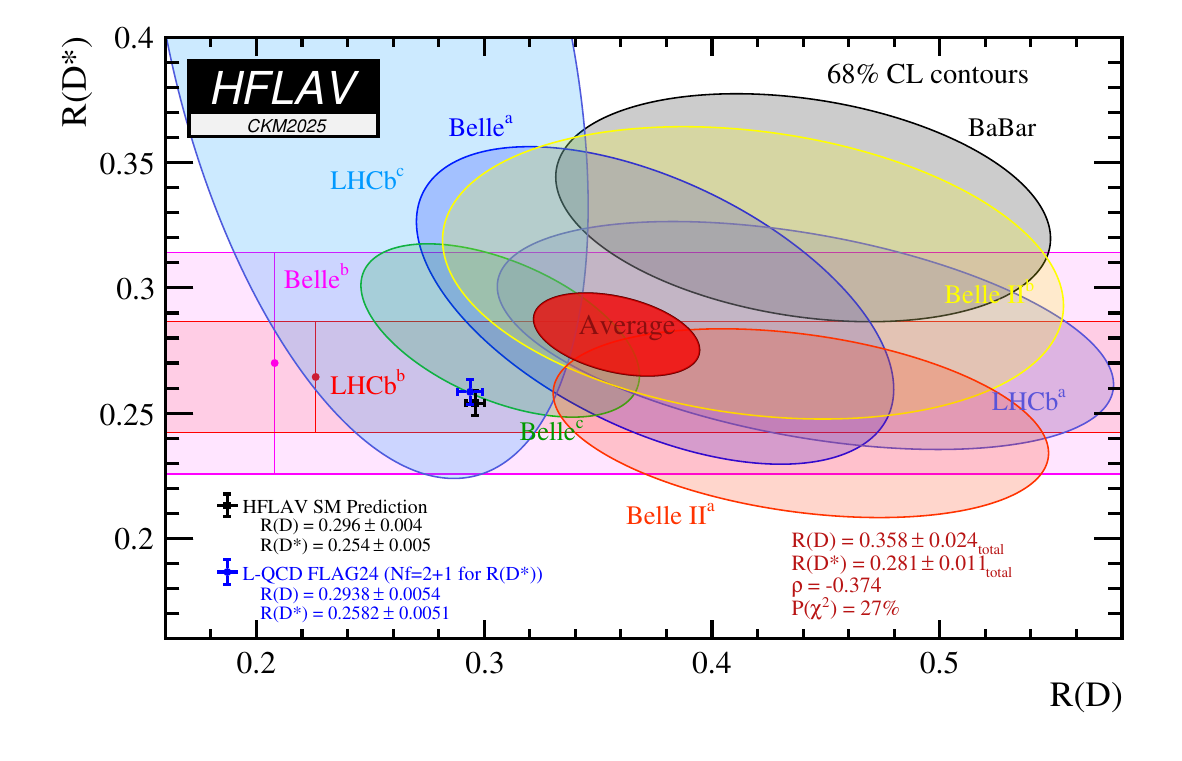}
    \includegraphics[width=0.4\textwidth]{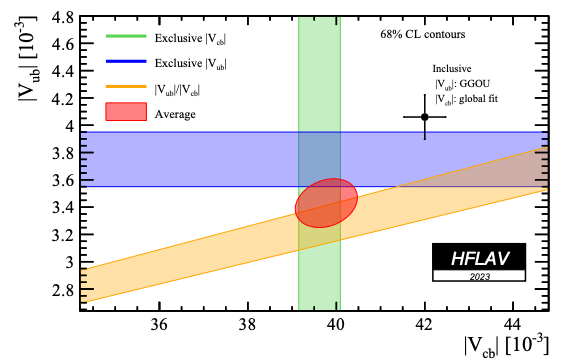}
    \caption{Summary of measurements of $\mathcal{R}(D)$ and $\mathcal{R}(D^{*})$ (left) and $V_{ub}$ and $V_{cb}$ (right) by the Heavy Flavor Averaging Group~\protect\cite{HFLAV23}.}
    \label{fig:RDstar_combined}
\end{figure}

The semileptonic $b$-hadron decays have a simple Standard Model (SM) description with a tree level diagram and high branching fractions. They are powerful probes for testing the SM and performing searches for physics beyond the Standard Model (BSM). In the SM the couplings of the gauge bosons are independent of the lepton flavour, also known as Lepton Flavour Universality (LFU). If any violation of LFU is measured it may lead to a clear sign of BSM physics. To test any discrepancies between decays to different leptons the ratio of branching fractions of $b$-hadrons having a $\mathrm{\tau}$ or a $\mathrm{\mu}$ lepton in their final states has been probed, defined as  $\mathcal{R}(H_c) = \frac{\mathcal{B}(H_b \rightarrow H_c \tau \nu)}{\mathcal{B} (H_b \rightarrow H_c \mu \nu)}$, where $H_c = D^*, D^+, J/\psi$ and $H_b = B^{\pm}, B^0$. Measurements of the ratios $\mathcal{R}(D)$ and $\mathcal{R}(D^*)$ have been performed at Belle, BaBar and LHCb with their average leading to an overall tension of 3.8 $\sigma$ with the SM prediction~\cite{HFLAV23}, shown in Fig.~\ref{fig:RDstar_combined} (left). Another intriguing discrepancy in these decays occurs in the determination of the CKM matrix elements, in particular $V_{ub}$ and $V_{cb}$ which can be performed using either inclusive or exclusive semileptonic $B$-meson decays. Inclusive determinations tend to yield higher values than exclusive ones, with discrepancies at the level of about $2$--$3\sigma$~\cite{HFLAV23} (see Fig.~\ref{fig:RDstar_combined} right). These tensions remain an open issue in flavour physics and may point to underestimated theoretical uncertainties or potential contributions from BSM physics. Various challenges arise when analysing these decays in a $pp$ environment. Firstly the inability of LHCb to detect neutrinos is a challenge for all semileptonic measurements performed there. Therefore specific procedures need to be followed to estimate the kinematic properties of the decay. Secondly, the events are polluted with various backgrounds that need careful modelling. In this contribution two recent measurements from LHCb are presented: a measurement of the branching fraction of $\Lambda \to p \mu^{-} \bar{\nu}_{\mu}$~\cite{LHCb:2025wld} and measurement of the form factor parameters using an angular analysis of the $B^0 \to D^{*-} \mu^{+} \nu_{\mu}$ decays~\cite{LHCb:2026amx}.


\section{Branching fraction measurement of $\Lambda \to p \mu^{-} \bar{\nu}_{\mu}$}

\begin{figure}[h]
    \centering
    \includegraphics[width=0.3\textwidth]{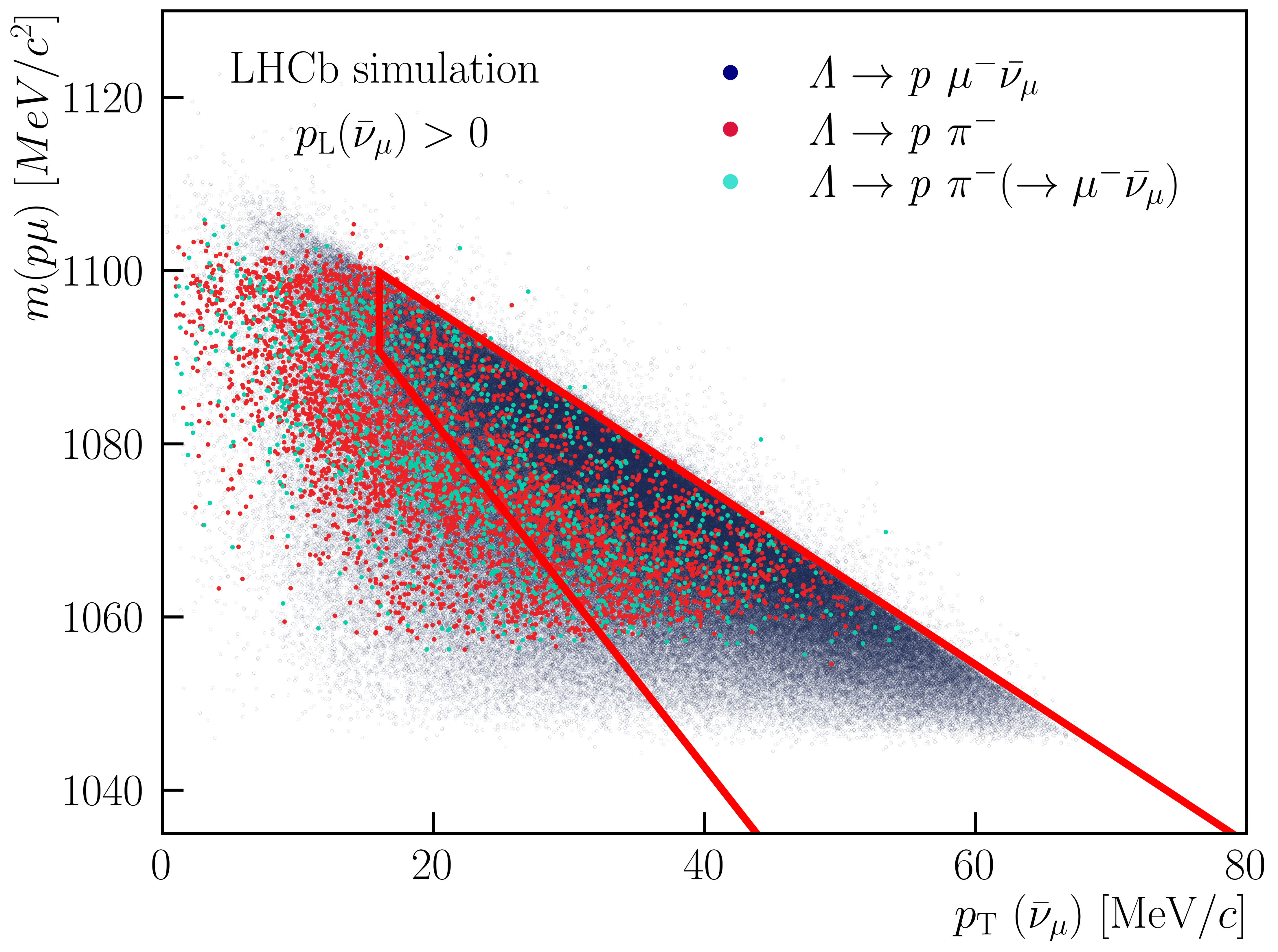}
    \includegraphics[width=0.3\textwidth]{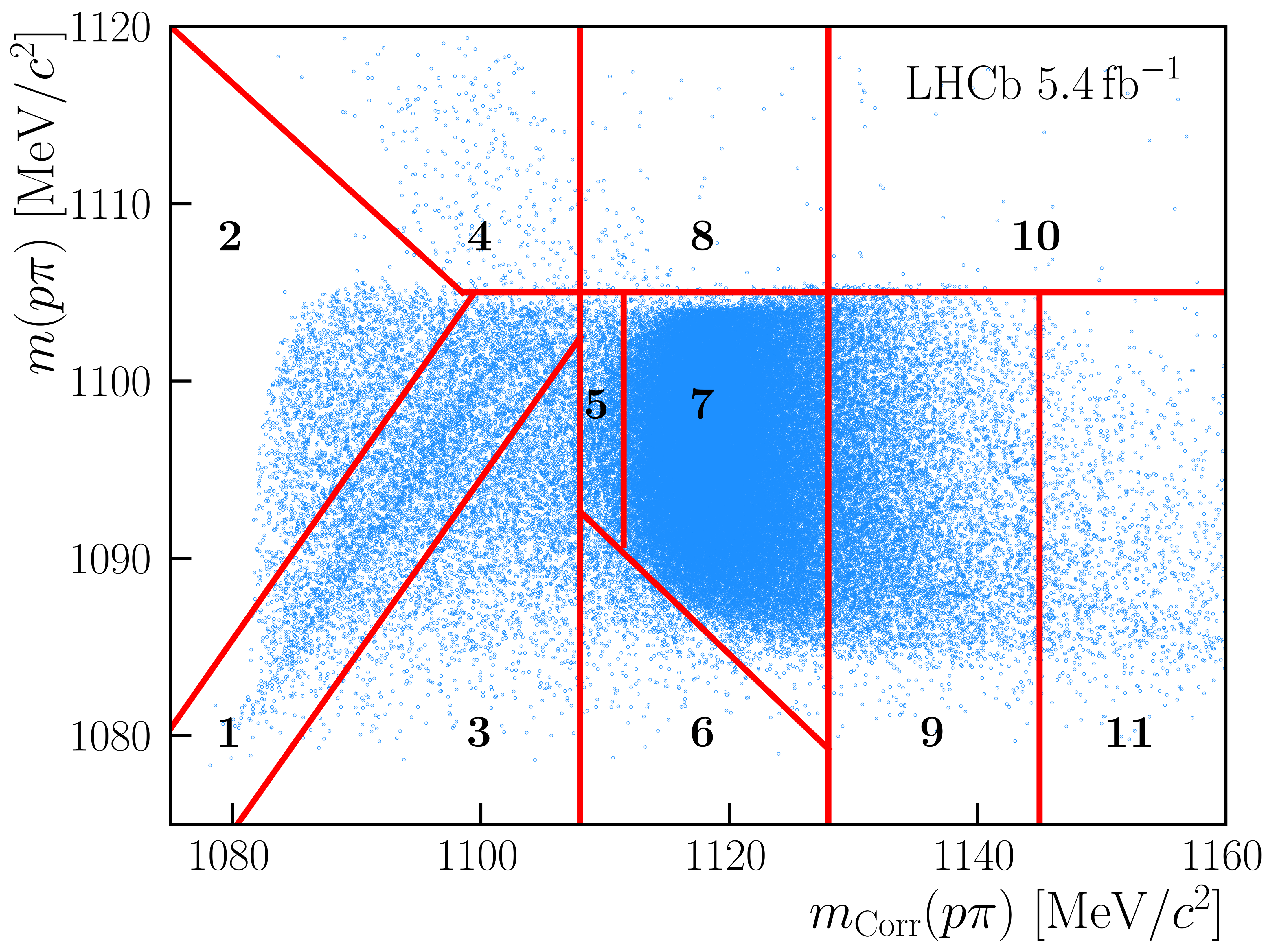}
    \caption{Graphical representation of the selection requirements in the (left) $p_T(\nu_\mu)$ vs.\ $m(p\mu)$ plane. Candidates inside the red box are selected. Right: Binning scheme of the $m_{\rm corr}(p\pi)$ vs.\ $m(p\pi)$ plane used to perform the 2D binned fit.}
    \label{fig:selection_Lambda}
\end{figure}

Semileptonic hyperon decays can be sensitive probes of LFU in $s \to u$  decays. The theory prediction for the LFU observable of the value of the muon-to-electron decay rate ratio
($R_{\mu e}$) is $0.153 \pm 0.008$~\cite{Chang:2014iba} for the $\Lambda \to p$ case. This prediction does not depend on form factor inputs and is theoretically clean, allowing for a straightforward comparison with experimental measurements. In addition, predictions are provided for the ratios $\Gamma_{p \ell \bar{\nu}_\ell} / |V_{us}|^2$ for both the electron and muon modes~\cite{Bacchio:2025auj}. This decay offers access to the $V_{us}$ CKM matrix element essential for testing the unitarity of the first row of the CKM matrix. When the $V_{us}$ measurements are combined with the precisely measured value of $V_{ud}$, the unitarity condition $|V_{ud}|^2 + |V_{us}|^2 + |V_{ub}|^2 = 1$ shows a deviation at the $2\sigma$ level~\cite{PhysRevD.110.030001}. This has motivated renewed interest in precision measurements of $V_{us}$. This measurement presents the result of the $\Lambda \to p \mu^{-} \bar{\nu}_{\mu}$ branching fraction using LHCb data recorded between 2016 and 2018 with an integrated luminosity of $5.4\,\mathrm{fb}^{-1}$.
It is performed relative to the $\Lambda \to p \pi^-$ decay, whose branching ratio has been measured as \mbox{$\mathcal{B}(\Lambda \to p \pi^-) = 0.641 \pm 0.005.$}~\cite{PhysRevD.110.030001}. The $\Lambda \to p \mu^{-} \bar{\nu}_{\mu}$ branching ratio is expressed as
\begin{equation}
\mathcal{B}(\Lambda \to p \mu^- \nu_\mu) 
= \mathcal{B}(\Lambda \to p \pi^-) 
\frac{\epsilon_{\Lambda \to p \pi}}{\epsilon_{\Lambda \to p \mu \nu_\mu}} 
\frac{N_{\Lambda \to p \mu \nu_\mu}}{N_{\Lambda \to p \pi}},
\label{eq:8}
\end{equation}
where $\epsilon_{\Lambda \to p \pi}$ and $\epsilon_{\Lambda \to p \mu \nu_\mu}$ are the selection efficiencies of the normalisation and signal channels, and $N_{\Lambda \to p \pi}$ and $N_{\Lambda \to p \mu \nu_\mu}$ are their corresponding yields. To extract the signal and normalisation yields, specific selection criteria have been developed, which includes the transverse momentum of the missing neutrino. The visible momentum can be decomposed as $\vec{p}\,'_{p\mu} = (\vec{p}_{p\mu} \cdot \hat{x}_u, \vec{p}_{p\mu} \cdot \hat{y}_u, \vec{p}_{p\mu} \cdot \hat{f}) = (p'_{p\mu,x}, p'_{p\mu,y}, p'_{p\mu,z}),$ where $p_T(\nu_\mu) = \sqrt{p'^2_{p\mu,x} + p'^2_{p\mu,y}}$. After computing $p_T(\nu_\mu)$ from the projections of the proton and muon momenta, a requirement of a positive argument inside the square root that  corresponds to the condition 
$p_T(\nu_\mu) < \frac{m_\Lambda^2 - m_{p\mu}^2}{2 m_\Lambda}$ is applied.
Signal decays satisfy this condition by construction, while combinatorial background typically does not. This imposes a constraint in the $p_T(\nu_\mu)$ vs.\ $m(p\mu)$ plane. A tighter selection is applied in this plane (see Fig.~\ref{fig:selection_Lambda} (left)). In addition, a requirement in the Armenteros--Podolanski~\cite{Podolanski01011954} plane is introduced, selecting candidates within an ellipse of the plane where the signal density is higher.


The normalisation yield is obtained from a fit to the $m(p\pi)$ distribution of candidates in data that pass the selection criteria for the normalisation channel. The signal yield is measured using a binned maximum-likelihood fit, following the procedure described in Ref.~\cite{LHCb:2025wld}, in bins of the $m_{\rm corr}(p\pi)$ (computed by scaling the total pion momentum so the $\Lambda$ reconstructed momentum points back to the PV) vs.\ $m(p\pi)$ plane, with the binning shown in Fig.~\ref{fig:selection_Lambda} (right).


The fit is performed using simulation templates for $\Lambda \to p \mu^{-} \bar{\nu}_{\mu}$, $\Lambda \to p \pi^-$, \mbox{$\Lambda \to p \pi^- (\to \mu^- \nu_\mu)$}, and the combinatorial background (modelled using minimum bias candidates). The measurement reports 
$$
\mathcal{B}(\Lambda \to p \mu^- \nu_\mu) = (1.462 \pm 0.016 \, (\rm stat) \pm 0.100 \, (\rm syst) \pm 0.011 \, (\rm norm)) \times 10^{-4} = (1.46 \pm 0.10) \times 10^{-4},
$$
corresponding to a total uncertainty of 6.9\%. This represents a factor of two improvement in precision compared to the previous best result obtained by BESIII~\cite{BESIII:2021ynj}. $|V_{us}|$ is extracted using lattice QCD predictions~\cite{Bacchio:2025auj}, yielding $|V_{us}| = 0.235 \pm 0.016$ with the more conservative input, and $|V_{us}| = 0.2459 \pm 0.0085$ with the alternative prediction featuring smaller uncertainties but less conservative assumptions. Using the precisely measured electron-mode branching fraction~\cite{PhysRevD.110.030001},
$
\mathcal{B}(\Lambda \to p e^{-} \bar{\nu}_{e}) = (8.34 \pm 0.14) \times 10^{-4}
$
the LFU ratio is determined as 
$
R_{\mu e} = 0.175 \pm 0.012.
$ 
This value is consistent with the main lattice QCD prediction of $0.1735 \pm 0.0098$~\cite{Bacchio:2025auj} at the $0.1\sigma$ level, with the more precise but less conservative lattice result of $0.16638 \pm 0.00020$~\cite{Bacchio:2025auj} at the $0.95\sigma$ level, and with the NLO prediction of $0.153 \pm 0.008$~\cite{LHCb:2025evf} at the $1.5\sigma$ level.

\section{$B^0 \to D^{*-} \mu^{+} \nu_{\mu}$ angular analysis}
differential observables of $B^0 \to D^{*-} \mu^{+} \nu_{\mu}$ decays (decay angles and squared momentum transferred to the lepton pair ($q^2$)) offer the opportunity to characterise the full decay rate and offer sensitivity to BSM models and hadronic dynamics of the decay. In this analysis the kinematic information of $B^0 \to D^{*-} \mu^{+} \nu_{\mu}$ decays is used to measure the hadronic form factors via a five-dimensional fit to the decay angles,  $q^2$, and the squared invariant mass missing from the visible system ($m_{miss}^2$). This is the first measurement of hadronic form factor parameters with $B^0 \to D^{*-} \mu^{+} \nu_{\mu}$ decays at LHCb using the Caprini--Lellouch--Neubert (CLN)~\cite{Caprini_1998}, Boyd--Grinstein--Lebed (BGL)~\cite{Boyd:1997kz}, and the Bernlochner--Ligeti--Papucci--Robinson (BLPR)~\cite{Bernlochner:2017jka} formalism. Data recorded by the LHCb detector within 2011–2012 are used corresponding to $pp$ collisions at centre-of-mass energies of 7 and 8 TeV, with integrated luminosity of 3.0 fb$^{-1}$. 


Due to the missing neutrino in the final state, the $B^0$ meson momentum can be determined up to a quadratic ambiguity where a solution for the $B^0$ momentum is selected randomly. This results in resolutions of 8 to 12$\%$ of the decay angles $\theta_\ell$, $\theta_d$ and $\chi$, and the $q^2$. \mbox{$m_{miss}^2 = (p_{B^{0}}-p_{D^*}-p_{\mu^+})^2$} is used to separate signal from background components, and is calculated using the rest-frame approximation~\cite{LHCb:2023zxo}. 

\begin{figure}[htbp] 
  \begin{center}
    \includegraphics[width=0.3\textwidth]{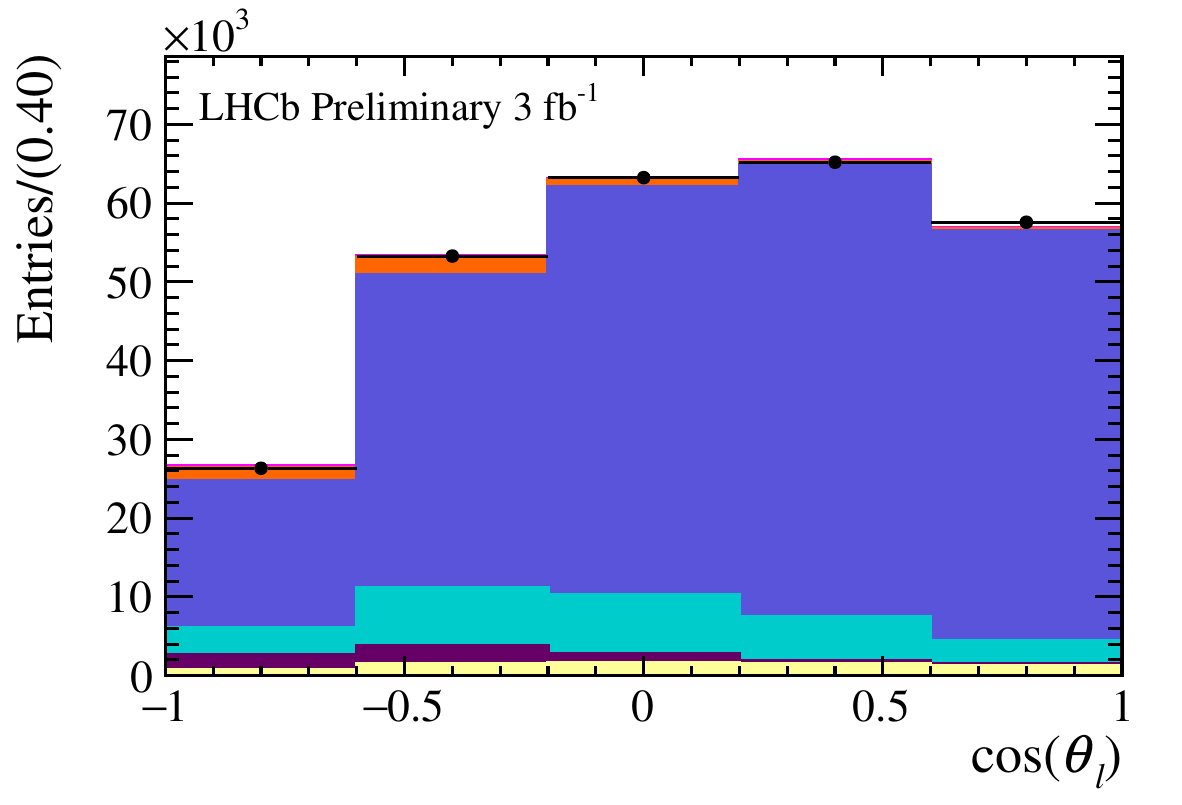}
    \includegraphics[width=0.3\textwidth]{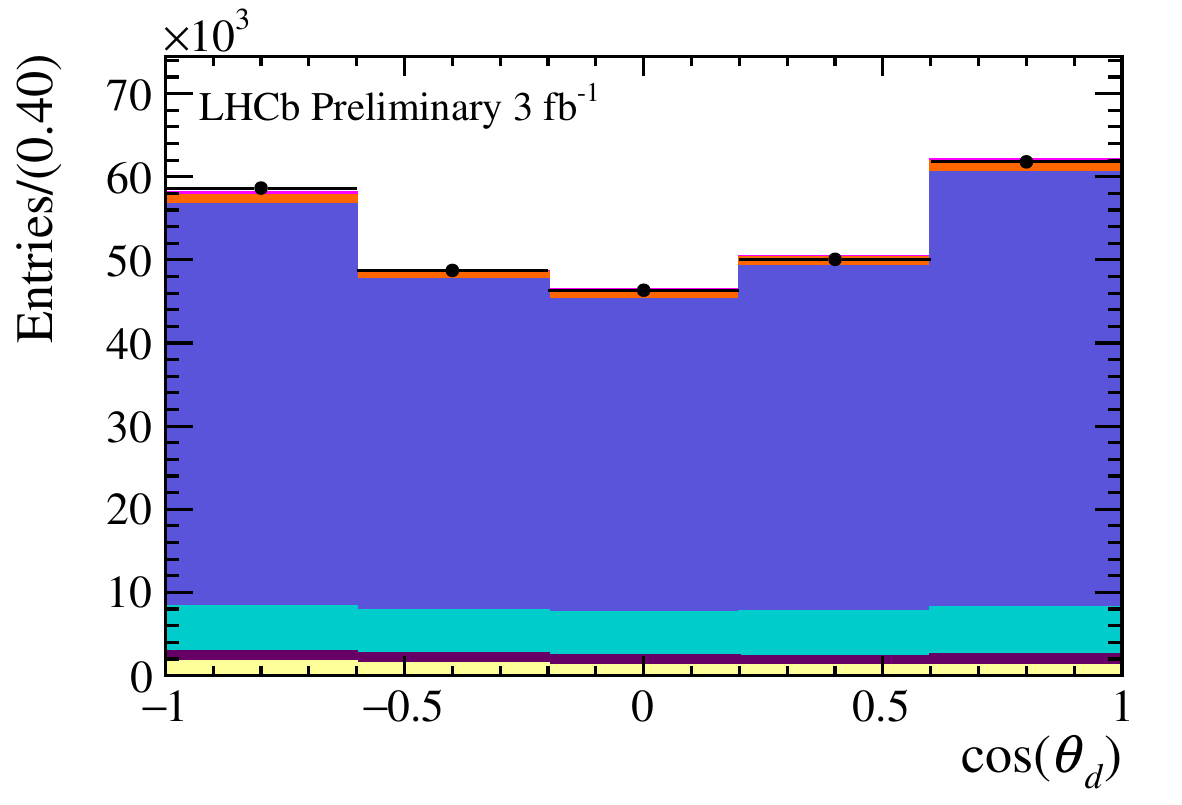}
    \includegraphics[width=0.3\textwidth]{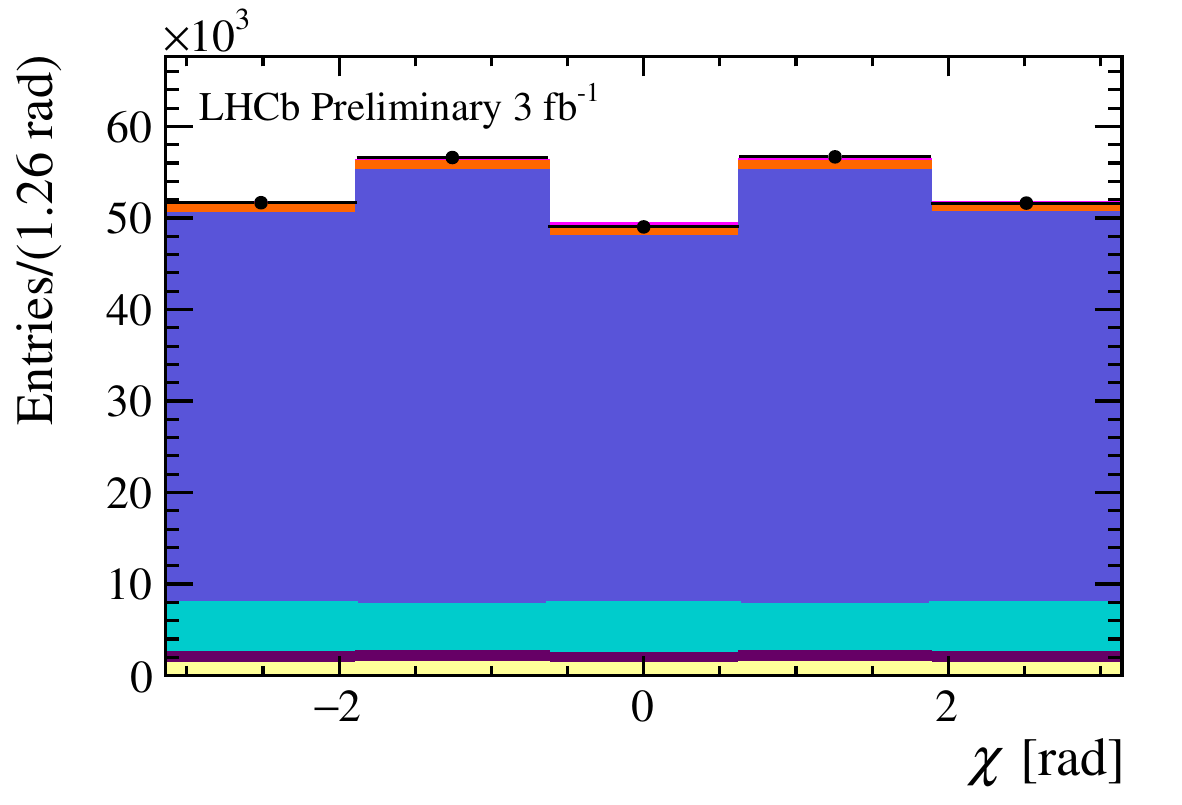}\\ 
    \includegraphics[width=0.3\textwidth]{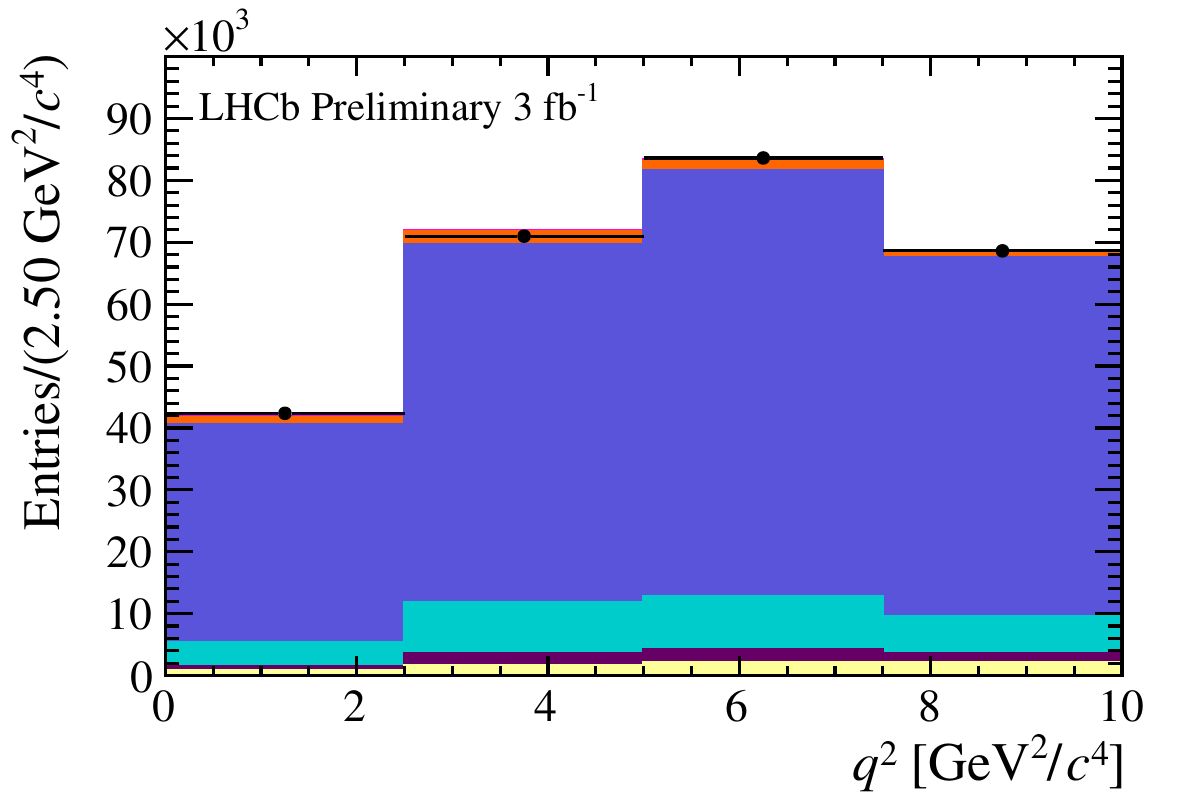}
    \includegraphics[width=0.3\textwidth]{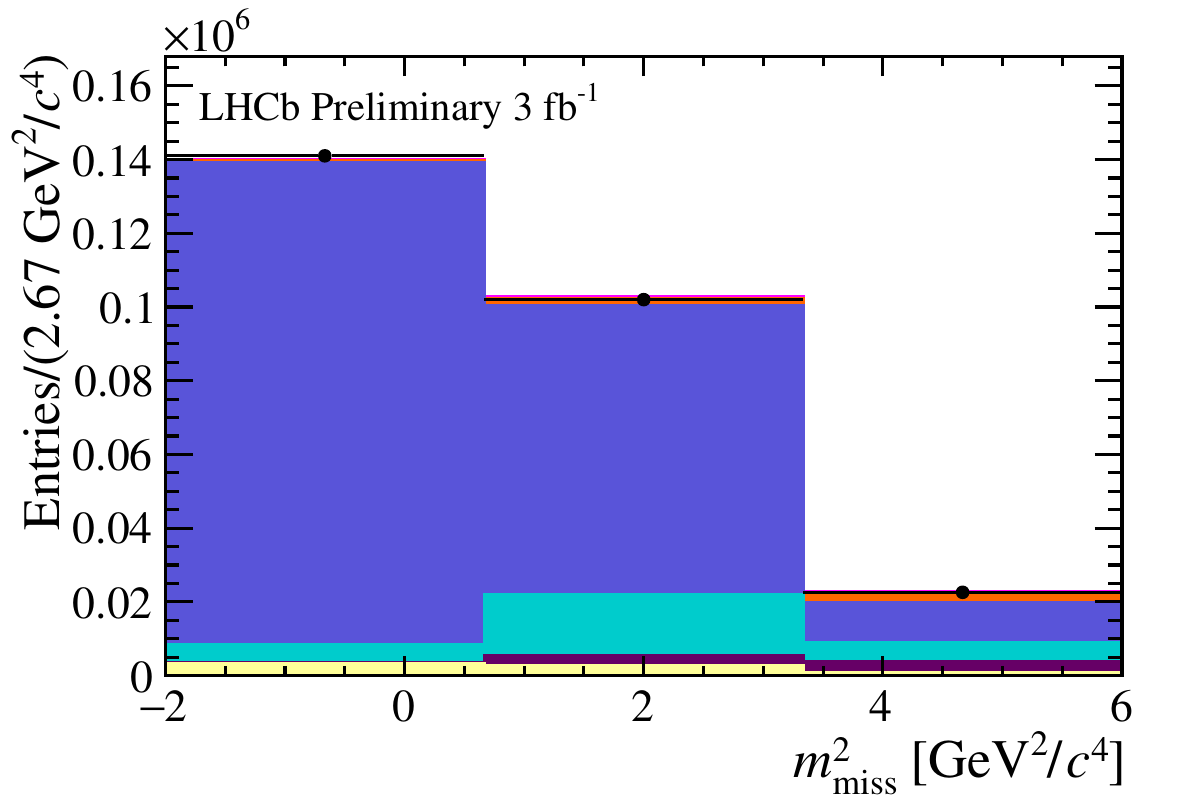}
    \includegraphics[width=0.3\textwidth]{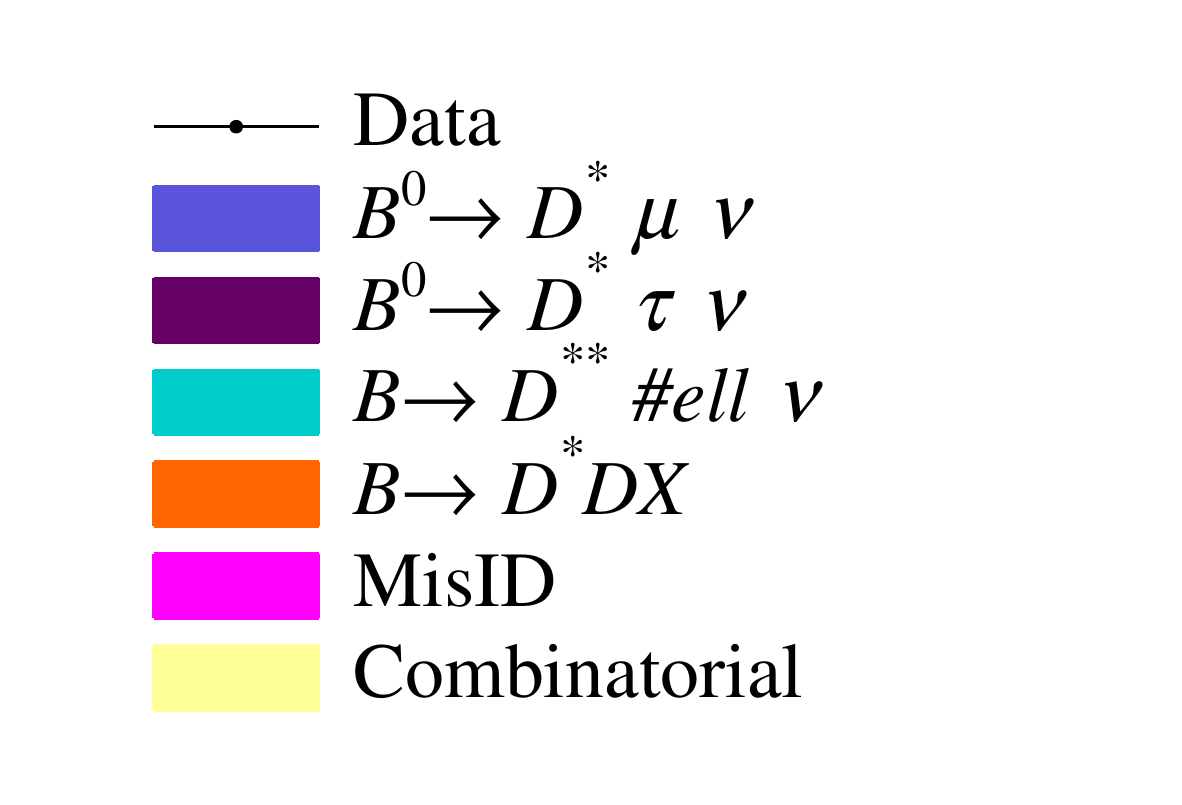} 
  \end{center} 
  \caption{Distributions of the decay angles, $q^2$ \ and $m_{miss}^2$ with the fit result also shown.}
  \label{fig:Fit}
\end{figure}

The five-dimensional histogram templates are constructed using 5 bins for the three angles, 4 bins in $q^2$ and 3 bins in $m_{miss}^2$. Each contribution to the fit model consists of a template histogram derived either from simulation or from specifically selected control samples in data. The hadronic form factor parameters in the signal model are varied with the {\sc HAMMER} tool~\cite{Bernlochner:2020tfi,hammer_zenodo}, using  the {\sc RooFit} interface~\cite{GarciaPardinas:2020yrd}. The most abundant background originates from $B\to D^{**}\mu\nu_{\mu}$ decays, including the lightest narrow $D^{**}$ states: $D_{1}(2420)$, $D^{*}_{2}(2460)$ and $D'_{1}(2430)$. Fits to the data are performed making assumptions on three  form factor parametrisations (CLN, BGL and BLPR) for the signal $B^0 \to D^{*-} \mu^{+} \nu_{\mu}$ decays where their hadronic form factor parameters are determined. The fit projections using the BGL parametrisation are shown in Fig.~\ref{fig:Fit}. The Bayesian information criterion (BIC) method~\cite{10.1214/aos/1176344136} has been used to choose the order of the BGL coefficient series. This consists of calculating the quantity $\mathrm{BIC} = \chi^2 + k \ln(n)$, where $\chi^2$ quantifies the fit quality, $k$ is the number of model parameters, and $n$ is the number of bins, for every possible combination of parameters. The configuration yielding the minimum BIC value is chosen as the nominal configuration. The parameters measured with this parametrisation are: $a_0$, $a_1$, $b_1$, $c_1$, $c_2$. The value of $b_0$ is set to be 0.013 as measured in the BGL$_{222}$ fit from Ref.~\cite{Bernlochner:2019ldg}. When fitting with CLN the $R_{0}$ parameter has been fixed to the value of 1.15 given the limited sensitivity of this measurement to the helicity-suppressed amplitude. Due to limited sensitivity to $\chi_3'$ in BLPR the value is fixed from prediction in Ref.~\cite{Bernlochner:2017jka}. The measured values for the hadronic form factor parameters are given in Table~\ref{tab:SMresults} together with their uncertainties and statistical correlations. The measured values of the BGL parameters satisfy the unitarity constraint. The largest systematic uncertainties arise from the limited simulation sample size, corrections of the simulation, truncation of BGL series and fixed form factor parameters.

\begin{table}[htbp]
\setlength{\tabcolsep}{3pt}
\centering
\caption{Form factor parameter fit results of CLN, BGL and BLPR. On the left measured values are reported alongside the statistical and systematic uncertainties. On the right the statistical correlation matrix is shown.}
\label{tab:SMresults}
\resizebox{0.6\textwidth}{!}{ 
\begin{tabular}{lccccccc}
\hline			
\multicolumn{7}{c} {CLN parametrisation} \\
\hline
    &  & \multicolumn{4}{c}{Statistical correlations} \\ 
    &  & &  $ R_1$ & $ R_2$ & $ \rho^2$ & & \\  
    $R_1$ & 1.303 $\pm$ 0.032 (stat) $\pm$ 0.049 (syst) & $  R_1$ & 1.00  & 0.66 & $-0.61$ & & \\ 
    $R_2$ & 0.859 $\pm$ 0.014 (stat) $\pm$ 0.031 (syst) & $ R_2$ &  & 1.00 & $-0.77$& & \\ 
    $\rho^2$ & 1.211 $\pm$ 0.020 (stat) $\pm$ 0.030 (syst) & $ \rho^2$ &  &  & \phantom{$-$}1.00 & & \\  
\hline
\multicolumn{7}{c} {BGL parametrisation}\\
\hline
&  & \multicolumn{4}{c}{Statistical correlations} \\ 
    &  & &  $ a_0$ & $  a_1$ & $ b_1$ & $ c_1$ &  $ c_2$\\ 
    $a_0$ & 0.026 $\pm$ 0.001 (stat) $\pm$ 0.002 (syst) & $ a_0$ &  1.00 & $-0.91$ & \phantom{$-$}0.50 & \phantom{$-$}0.35 & $-0.27$\\ 
    $a_1$ & $-0.039$ $\pm$ 0.022 (stat) $\pm$ 0.031 (syst) & $ a_1$ &  & \phantom{$-$}1.00 & $-0.56$ & $-0.23$ & \phantom{$-$}0.15 \\ 
    $b_1$ & -0.008 $\pm$ 0.004 (stat) $\pm$ 0.031 (syst) & $ b_1$ &  &  & \phantom{$-$}1.00 & \phantom{$-$}0.58 & $-0.48$\\  
    $c_1$ & $-0.001$ $\pm$ 0.001 (stat) $\pm$ 0.007 (syst) & $ c_1$ & &  & & \phantom{$-$}1.00& $-0.97$\\  
    $c_2$ & 0.019 $\pm$ 0.026 (stat) $\pm$ 0.026 (syst) & $ c_2$ &  &  &  &  & \phantom{$-$}1.00\\  
\hline
\multicolumn{7}{c} {BLPR parametrisation}\\
\hline
&  &  \multicolumn{4}{c}{Statistical correlations}\\ 
    &  & &  $ \bar{\rho}_*^2$ & $ \chi_2(1)$ & $ \chi_2^{'}(1)$ & $ \eta(1)$ &  $\eta(1)'$\\ 
   $\bar{\rho}_*^2$ & 1.28 $\pm$ 0.03 (stat) $\pm$ 0.04  (syst)  & $\bar{\rho}_*^2$ & 1.00 & $-0.28$ & $-0.68$ & $-0.27$ & \phantom{$-$}0.83 \\ 
    $\chi_2(1)$ & $-0.11$ $\pm$ 0.12 (stat) $\pm$ 0.29 (syst) &  $\chi_2(1)$ & & \phantom{$-$}1.00 & \phantom{$-$}0.11 & $-0.68$ & $-0.33$\\ 
    $\chi_2^{'}(1)$ & 0.38 $\pm$ 0.18 (stat) $\pm$ 0.20 (syst) &  $\chi_2^{'}(1)$ & & & \phantom{$-$}1.00 & \phantom{$-$}0.32 & $-0.85$\\ 
    $\eta(1)$ & 0.43 $\pm$ 0.14 (stat) $\pm$ 0.36  (syst) &  $\eta(1)$  &  &  & & \phantom{$-$}1.00 & $-0.38$ \\  
    $\eta(1)'$ & $-0.52$ $\pm$ 0.52 (stat) $\pm$ 0.47 (syst) &  $\eta(1)'$ &  &  &  & & \phantom{$-$}1.00 \\ 
\hline
\end{tabular}}
\end{table}

\begin{figure}[htbp] 
  \begin{center}
    \includegraphics[width=0.3\textwidth]{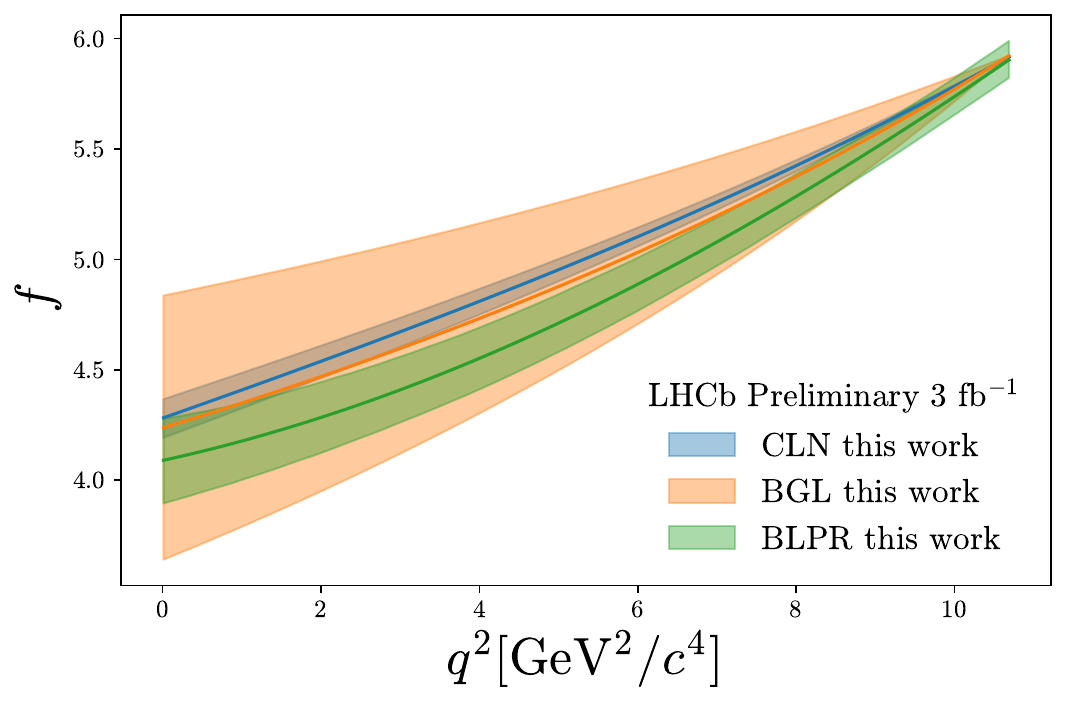}
    \includegraphics[width=0.3\textwidth]{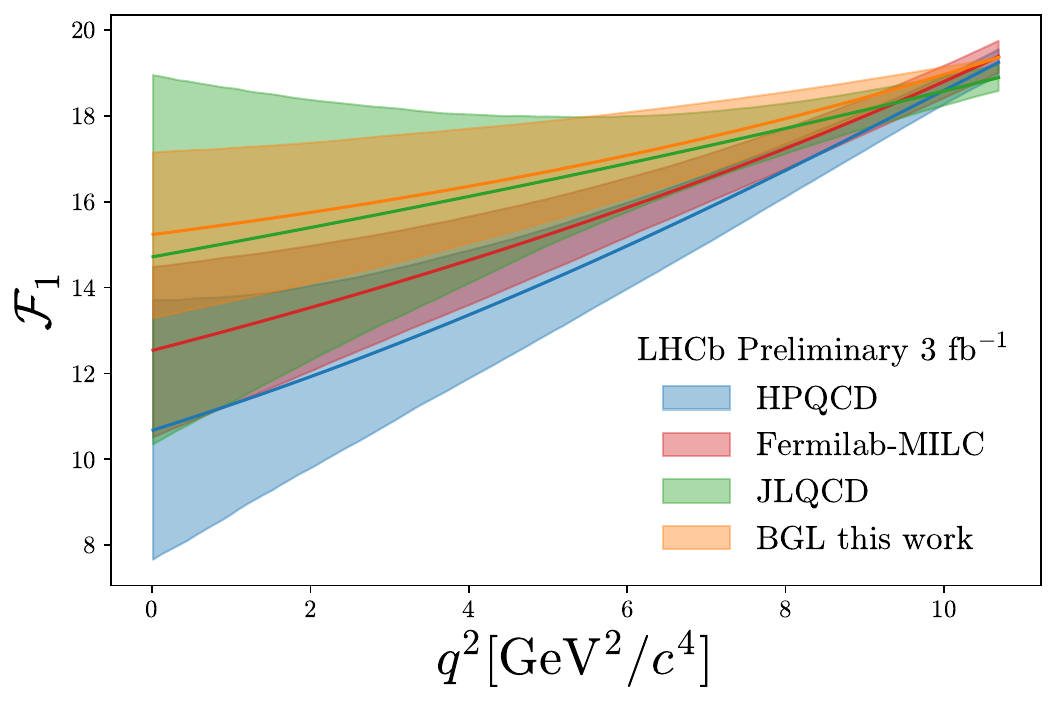}
    \includegraphics[width=0.3\textwidth]{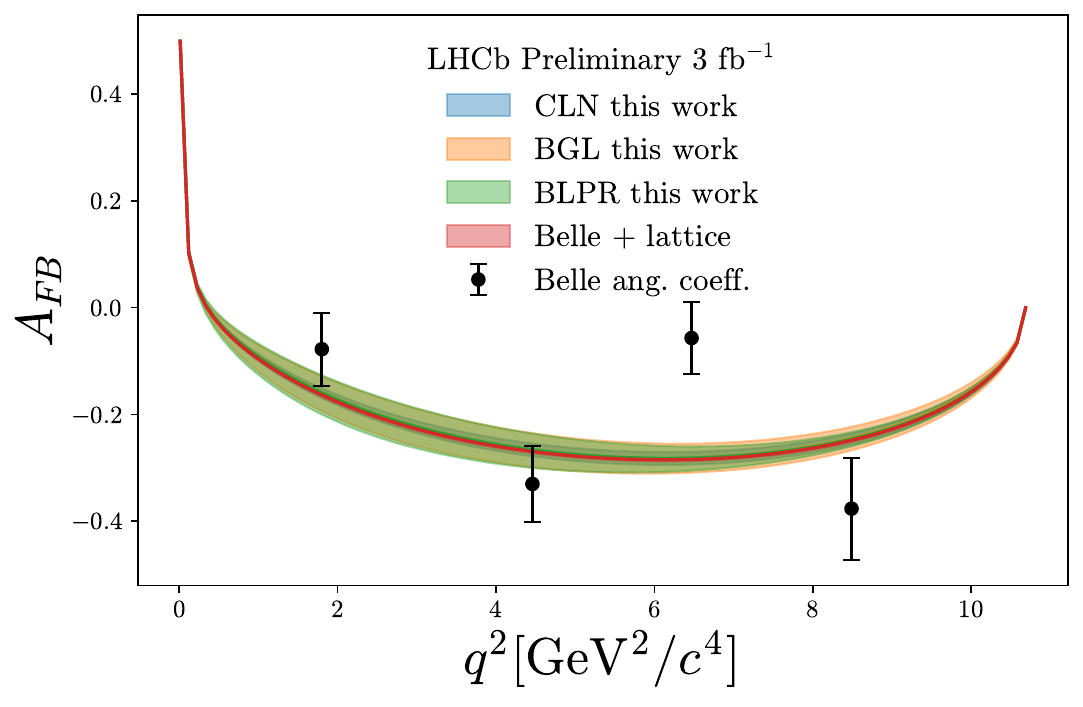}
  \end{center} 
  \caption{Left: Comparison of the $f$ results obtained in this form factor measurement using CLN (blue), BGL (orange), and BLPR (green) parametrisations in the BGL basis. The uncertainty bands assume uncorrelated systematic uncertainties. Middle: Comparison of the results obtained for the $\mathcal{F}_{1}$ form factor using the BGL to lattice QCD calculations from HPQCD (blue), Fermilab-MILC (red) and JLQCD (green). The uncertainty bands for this work are derived assuming that systematic uncertainties are uncorrelated. Right: Distribution of forward-backward asymmetry of the muon ($A_{\rm FB}$) compared with Belle results from Ref.~\protect\cite{Bernlochner:2017jka} (red) and Ref.~\protect\cite{Belle:2023xgj} (black points).}
  \label{fig:formfactor_parametrisations}
\end{figure}


Fig.~\ref{fig:formfactor_parametrisations} (left) shows a comparison of the results obtained in this measurement in the BGL basis. A tension is seen between the CLN and BLPR $f(q^2)$ form factor that can be expected given the restrictive CLN assumptions. Fig.~\ref{fig:formfactor_parametrisations} (middle) compares $\mathcal{F}_1$ results obtained using the BGL parametrisation to the lattice QCD calculations by the HPQCD~\cite{Harrison:2023dzh}, the Fermilab-MILC ~\cite{FermilabLattice:2021cdg} and the JLQCD collaboration~\cite{Aoki:2023qpa}. The form factors $f(q^2)$ and $g(q^2)$ are compatible while in $\mathcal{F}_1(q^2)$ the results agree better with Fermilab-MILC and JLQCD than HPQCD. Similar can be seen for CLN and BLPR from Ref~\cite{LHCb:2026amx}. In Fig.~\ref{fig:formfactor_parametrisations} (right) the muon forward-backward asymmetry, $A_{\rm FB}$ is calculated using the hadronic form factor parameter results. It shows excellent agreement with the results from the BLPR fit~\cite{Bernlochner:2017jka} and as well the Belle measurement of the angular coefficients~\cite{Belle:2023xgj}. This is the first measurement of the hadronic form factor parameters in $B^0 \to D^{*-} \mu^{+} \nu_{\mu}$ decays, based on a five-dimensional binned likelihood fit of LHCb data. The results of the form factor parameters from BGL, BLPR, and CLN give mutually consistent results and are compatible with lattice calculations and theoretical predictions, while achieving improved precision. 



\section{Summary and prospects}




These proceedings present the worlds best measurement of the branching fraction of $\Lambda \to p \mu^{-} \bar{\nu}_{\mu}$ together with the first LHCb angular analysis in $B^0 \to D^{*-} \mu^{+} \nu_{\mu}$ measuring form factor parameters in CLN, BGL and BLPR. The measurements are consistent with lattice and theory determinations. form factor modelling introduces a significant systematic uncertainty and this result helps mitigate that issue by providing inputs derived directly from LHCb data. 
Within the next planned upgrades of LHCb higher instantaneous luminosities will allow to collect large datasets improving the statistical precision of measurements. The expected precision on semileptonic measurements over the years is estimated at around 3 $\%$ on the ratios of branching fractions~\cite{ATLAS:2025lrr} if in the years to come all the different challenges in computation, background modeling and estimation are met. Overall, the LHCb programme of $b \to c \ell \nu$ measurements plays a central role in flavour physics, offering complementary information to $B$-factory experiments and continuing to probe possible deviations from Standard Model expectations.

\section*{Acknowledgments}

B. M. acknowledges support by UK Science and Technology Facilities Council and the University of Manchester.

\section*{References}
\bibliography{moriond}

@article{HFLAV23,
  author        = "Banerjee, S. and others",
  collaboration = "Heavy Flavor Averaging Group",
  title         = {Averages of b-hadron, c-hadron, and tau-lepton properties as of 2023},
  eprint        = "2411.18639",
  archivePrefix = "arXiv",
  primaryClass  = "hep-ex",
  year          = "2026",
  journal       = {Phys. Rev. D},
  volume        = {113},
  pages         = {012008},
}

@article{LHCb:2023zxo,
    author = "Aaij, Roel and others",
    collaboration = "LHCb",
    title = "{Measurement of the ratios of branching fractions $\mathcal{R}(D^{*})$ and $\mathcal{R}(D^{0})$}",
    eprint = "2302.02886",
    archivePrefix = "arXiv",
    primaryClass = "hep-ex",
    reportNumber = "LHCb-PAPER-2022-039, CERN-EP-2022-284",
    doi = "10.1103/PhysRevLett.131.111802",
    journal = "Phys. Rev. Lett.",
    volume = "131",
    pages = "111802",
    year = "2023"
}

@article{Chang:2014iba,
    author = "Chang, Hsi-Ming and Gonz{\'a}lez-Alonso, Martin and Martin Camalich, Jorge",
    title = "{Nonstandard Semileptonic Hyperon Decays}",
    eprint = "1412.8484",
    archivePrefix = "arXiv",
    primaryClass = "hep-ph",
    reportNumber = "UCSD-PTH-14-11",
    doi = "10.1103/PhysRevLett.114.161802",
    journal = "Phys. Rev. Lett.",
    volume = "114",
    number = "16",
    pages = "161802",
    year = "2015"
}

@article{Bacchio:2025auj,
    author = "Bacchio, Simone and Konstantinou, Andreas",
    title = "{Study of the {\ensuremath{\Lambda}}{\textrightarrow}p{\ensuremath{\ell}}{\ensuremath{\nu}}{\textasciimacron}{\ensuremath{\ell}} Semileptonic Decay in Lattice QCD}",
    eprint = "2507.09970",
    archivePrefix = "arXiv",
    primaryClass = "hep-lat",
    doi = "10.1103/bvs9-wcsj",
    journal = "Phys. Rev. Lett.",
    volume = "135",
    number = "23",
    pages = "231901",
    year = "2025"
}

@article{PhysRevD.110.030001,
  title = {Review of Particle Physics},
  author = {Navas, S. and others},
  collaboration = {Particle Data Group Collaboration},
  journal = {Phys. Rev. D},
  volume = {110},
  issue = {3},
  pages = {030001},
  numpages = {5},
  year = {2024},
  month = {Aug},
  publisher = {American Physical Society},
  doi = {10.1103/PhysRevD.110.030001},
  url = {https://link.aps.org/doi/10.1103/PhysRevD.110.030001}
}

@article{Podolanski01011954,
author = {J. Podolanski and R. Armenteros},
title = {III. Analysis of V-events},
journal = {The London, Edinburgh, and Dublin Philosophical Magazine and Journal of Science},
volume = {45},
number = {360},
pages = {13--30},
year = {1954},
publisher = {Taylor \& Francis},
doi = {10.1080/14786440108520416},
URL = { 
        https://doi.org/10.1080/14786440108520416
},
eprint = { 
        https://doi.org/10.1080/14786440108520416
}
}

@article{BESIII:2021ynj,
    author = "Ablikim, M. and others",
    collaboration = "BESIII",
    title = "{First Measurement of the Absolute Branching Fraction of $\Lambda \to p \mu^- \bar{\nu}_{\mu}$}",
    eprint = "2107.06704",
    archivePrefix = "arXiv",
    primaryClass = "hep-ex",
    doi = "10.1103/PhysRevLett.127.121802",
    journal = "Phys. Rev. Lett.",
    volume = "127",
    number = "12",
    pages = "121802",
    year = "2021"
}

@article{LHCb:2025evf,
  author = "Aaij, Roel and others",
  collaboration = "LHCb",
  title = "{Observation of the very rare $\Sigma^+ \to p \mu^+ \mu^-$ decay}",
  eprint = "2504.06096",
  archivePrefix = "arXiv",
  primaryClass = "hep-ex",
  reportNumber = "LHCb-PAPER-2025-002, CERN-EP-2025-074",
  doi = "10.1103/r3v2-kmmp",
  journal = "Phys. Rev. Lett.",
  volume = "135",
  number = "5",
  pages = "051801",
  year = "2025"
}

@article{Caprini_1998,
title = {Dispersive bounds on the shape of {$B \to D^{*} \ell \nu$} form factors},
journal = {Nucl. Phys.},
volume = {B530},
number = {1},
pages = {153-181},
year = {1998},
issn = {0550-3213},
eprint         = "hep-ph/9712417v1",
archivePrefix  = "arXiv",
doi = {https://doi.org/10.1016/S0550-3213(98)00350-2},
url = {https://www.sciencedirect.com/science/article/pii/S0550321398003502},
author = {Caprini, I. and Lellouch, L. and Neubert, M.}
}

@article{Boyd:1997kz,
    archiveprefix = {arXiv},
    author = {Boyd, C. Glenn and Grinstein, Benjamin and Lebed, Richard F.},
    doi = {10.1103/PhysRevD.56.6895},
    eprint = {hep-ph/9705252},
    journal = {Phys. Rev.},
    volume = {D56},
    pages = {6895-6911},
    primaryclass = {hep-ph},
    reportnumber = {CMU-HEP-97-07A, UCSD-PTH-97-12},
    title = {{Precision corrections to dispersive bounds on form-factors}},
    year = {1997}
}

@article{Bernlochner:2017jka,
    author = "Bernlochner, F. U. and Ligeti, Z. and Papucci, M. and Robinson, D. J.",
    title = "{Combined analysis of semileptonic $B$ decays to $D$ and $D^*$: $R(D^{(*)})$, $|V_{cb}|$, and new physics}",
    eprint = "1703.05330",
    archivePrefix = "arXiv",
    primaryClass = "hep-ph",
    doi = "10.1103/PhysRevD.95.115008",
    journal = "Phys. Rev.",
    volume = "D95",
    number = "11",
    pages = "115008",
    year = "2017",
    extraPrefix    = "Erratum",
    extraVolume    = "97",
    extraPages     = "059902",
    extraYear      = "2018",
    extraDoi       = "10.1103/PhysRevD.95.115008",
}

@article{Bernlochner:2020tfi,
    author = "Bernlochner, F. U. and Duell, Stephan and Ligeti, Zoltan and Papucci, Michele and Robinson, Dean J.",
    title = "{Das ist der HAMMER: Consistent new physics interpretations of semileptonic decays}",
    eprint = "2002.00020",
    archivePrefix = "arXiv",
    primaryClass = "hep-ph",
    doi = "10.1140/epjc/s10052-020-8304-0",
    journal = "Eur. Phys. J.",
    volume = "C80",
    number = "9",
    pages = "883",
    year = "2020"
}

@misc{hammer_zenodo,
  author       = {Florian U. Bernlochner and Stephan Duell and Zoltan Ligeti and Michele Papucci and Dean J. Robinson},
  title        = "{HAMMER: Helicity amplitude module for matrix element reweighting}",
  version      = {1.4.1},
  year         = {2024},
  publisher    = {Zenodo},
  doi          = {10.5281/zenodo.11245573},
  url          = {https://doi.org/10.5281/zenodo.11245573}
}

@article{GarciaPardinas:2020yrd,
    author = "Garc\'\i{}a Pardi\~nas, J. and Meloni, S. and Grillo, L. and Owen, P. and Calvi, M. and Serra, N.",
    title = "{RooHammerModel: Interfacing the HAMMER software tool with HistFactory and RooFit}",
    eprint = "2007.12605",
    archivePrefix = "arXiv",
    primaryClass = "hep-ph",
    doi = "10.1088/1748-0221/17/04/T04006",
    journal = "JINST",
    volume = "17",
    number = "04",
    pages = "T04006",
    year = "2022"
}

@article{10.1214/aos/1176344136,
author = {Gideon Schwarz},
title = "{Estimating the dimension of a model}",
volume = {6},
journal = {Ann. Statist.},
number = {2},
publisher = {Institute of Mathematical Statistics},
pages = {461 -- 464},
year = {1978},
doi = {10.1214/aos/1176344136},
URL = {https://doi.org/10.1214/aos/1176344136}
}

@article{Bernlochner:2019ldg,
    author = "Bernlochner, Florian U. and Ligeti, Zoltan and Robinson, Dean J.",
    title = "{N = 5, 6, 7, 8: Nested hypothesis tests and truncation dependence of $|V_{cb}|$}",
    eprint = "1902.09553",
    archivePrefix = "arXiv",
    primaryClass = "hep-ph",
    doi = "10.1103/PhysRevD.100.013005",
    journal = "Phys. Rev.",
    volume = "D100",
    number = "1",
    pages = "013005",
    year = "2019"
}

@article{Harrison:2023dzh,
    author = "Harrison, J. and Davies, C. T. H.",
    collaboration = "HPQCD",
    title = "{${B \to D^{*}}$ and ${B_s \to D_s^{*}}$ vector, axial-vector and tensor form factors for the full $q^2$ range from lattice QCD}",
    eprint = "2304.03137",
    archivePrefix = "arXiv",
    primaryClass = "hep-lat",
    doi = "10.1103/PhysRevD.109.094515",
    journal = "Phys. Rev.",
    volume = "D109",
    number = "9",
    pages = "094515",
    year = "2024"
}

@article{FermilabLattice:2021cdg,
    author         = "Bazavov, A. and others",
    collaboration  = "Fermilab Lattice, MILC collaborations",
    title          = "{Semileptonic form factors for ${B \to D^* \ell \nu}$ at nonzero recoil from $2+1$-flavor lattice QCD}",
    eprint         = "2105.14019",
    archivePrefix  = "arXiv",
    primaryClass   = "hep-lat",
    reportNumber   = "FERMILAB-PUB-21-261-T, FERMILAB-PUB-21/261-T",
    doi            = "10.1140/epjc/s10052-022-10984-9",
    journal        = "Eur. Phys. J.",
    volume         = "C82",
    number         = "12",
    pages          = "1141",
    year           = "2022",
    extraPrefix    = "Erratum",
    extraVolume    = "83",
    extraPages     = "21",
    extraYear      = "2023",
    extraDoi       = "10.1140/epjc/s10052-023-00210-2"
}

@article{Aoki:2023qpa,
    author = "Aoki, Y. and Colquhoun, B. and Fukaya, H. and Hashimoto, S. and Kaneko, T. and Kellermann, R. and Koponen, J. and Kou, E.",
    collaboration = "JLQCD",
    title = {{$B \to D^*\ell \nu_{\ell}$ semileptonic form factors from lattice QCD with M{\"o}bius domain-wall quarks}},
    eprint = "2306.05657",
    archivePrefix = "arXiv",
    primaryClass = "hep-lat",
    reportNumber = "KEK-CP-393, OU-HET-1186",
    doi = "10.1103/PhysRevD.109.074503",
    journal = "Phys. Rev.",
    volume = "D109",
    number = "7",
    pages = "074503",
    year = "2024"
}

@misc{ATLAS:2025lrr,
    author = "ATLAS and Belle-II and CMS and LHCb",
    title = "{Projections for Key Measurements in Heavy Flavour Physics}",
    year = "2025",
    month = "3",
    note = "arXiv:2503.24346 "
}

@article{Belle:2023xgj,
    author = "Prim, M. T. and others",
    collaboration = "Belle",
    title = "{Measurement of angular coefficients of {$B \to D^{*} \ell \nu$}: Implications for {$|V_{cb}|$} and tests of lepton flavor universality}",
    eprint = "2310.20286",
    archivePrefix = "arXiv",
    primaryClass = "hep-ex",
    reportNumber = "Belle Preprint 2023-18; KEK Preprint 2023-32",
    doi = "10.1103/PhysRevLett.133.131801",
    journal = "Phys. Rev. Lett.",
    volume = "133",
    number = "13",
    pages = "131801",
    year = "2024"
}

@misc{LHCb:2025wld,
  author = "Aaij, Roel and others",
  collaboration = "LHCb",
  title = "{Branching fraction measurement of the $\Lambda \to p \mu^- \overline{\nu}_\mu$ decay}",
  year = "2025",
  eprint = "2511.15681",
  archivePrefix = "arXiv",
  primaryClass = "hep-ex",
  note = "arXiv:2511.15681"
}

@misc{LHCb:2026amx,
    author = "LHCb collaboration",
    title = "{Measurement of hadronic form-factor parameters with an angular analysis of $B^0 \rightarrow D^{*-} \mu^{+} \nu_{\mu}$ decays}",
    year = "2026",
    note = "LHCb-CONF-2026-001, CERN-LHCb-CONF-2026-001"
}

\end{document}